\documentclass[10pt, a4paper]{article}
\usepackage[utf8]{inputenc}
\usepackage{dcolumn,lscape}
\usepackage{amsmath,multicol,dcolumn,graphicx,amssymb}
\usepackage{exscale,amsthm,rotating}
\usepackage[numbers]{natbib}
\usepackage{hyperref}
\hypersetup{
  colorlinks   = true, 
  urlcolor     = blue, 
  linkcolor    = blue, 
  citecolor   = red 
}

\usepackage{tikz,dsfont,float}
\usepackage{caption}
\usepackage{subcaption}
\usepackage{tabularx}
\usepackage{multirow}
\usepackage{authblk}
\usepackage{booktabs}

\newcommand*{\thead}[1]{\multicolumn{1}{c}{\textbf{#1}}}

\title{Agrimonia: a dataset on livestock, meteorology and air quality in the Lombardy region, Italy}

\author[1,*]{Alessandro Fassò}
\author[1]{Jacopo Rodeschini}
\author[2]{Alessandro Fusta Moro}
\author[3]{Qendrim Shaboviq}
\author[4,5]{Paolo Maranzano}
\author[1]{Michela Cameletti}
\author[1]{Francesco Finazzi}
\author[2]{Natalia Golini}
\author[2]{Rosaria Ignaccolo}
\author[3]{Philipp Otto}

\affil[1]{University of Bergamo, Dept. of Economics, Via dei Caniana 2, 24127, Bergamo, Italy}
\affil[2]{University of Torino, Dept. of Economics and Statistics, Lungo Dora Siena 100A, 10153, Torino, Italy}
\affil[3]{Leibniz University Hannover, Institute of Cartography and Geoinformatics, Appelstrasse 9a, 30167, Hannover, Germany}
\affil[4]{University of Milano-Bicocca, Dept. of Economics, Management and Statistics, Piazza dell’Ateneo Nuovo 1, 20126, Milano, Italy}
\affil[5]{Fondazione Eni Enrico Mattei (FEEM), Corso Magenta 63, 20123, Milano, Italia}

\begin{document}

\maketitle

\begin{abstract}
The air in the Lombardy region, Italy, is one of the most polluted in Europe because of limited air circulation and high emission levels. There is a large scientific consensus that the agricultural sector has a significant impact on air quality. To support studies quantifying the role of the agricultural and livestock sectors on the Lombardy air quality, this paper presents a harmonised dataset containing daily values of air quality, weather, emissions, livestock, and land and soil use in the years 2016 - 2021, for the Lombardy region. The pollutant data come from the European Environmental Agency and the Lombardy Regional Environment Protection Agency, weather and emissions data from the European Copernicus programme, livestock data from the Italian zootechnical registry, and land and soil use data from the CORINE Land Cover project. The resulting dataset is designed to be used as is by those using air quality data for research.
\end{abstract}

\noindent%

\newpage


\section*{Background \& Summary} 

\begin{figure}[ht]
\centering
\includegraphics[width=\linewidth]{ 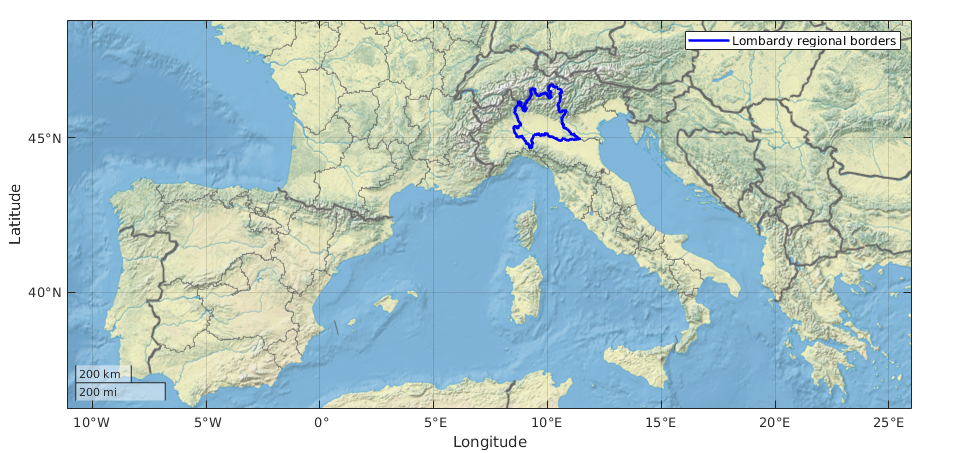}
\caption{The localisation of the Lombardy region in Northern Italy surrounded by the Alps. \label{fig:lombardy_italy}}
\end{figure}

Air pollutants may be categorised as primary or secondary. Primary pollutants are directly emitted to the atmosphere, whereas secondary pollutants are formed in the atmosphere from precursor gases through chemical reactions and microphysical processes. One of the key precursor gases for secondary particulate matter (PM) is ammonia (NH$_3$). This holds true for both large PM with aerodynamic diameter less then $10 \mu m$ (PM$_{10}$) and fine PM with aerodynamic diameter less then $2.5 \mu m$ (PM$_{2.5}$). There is a large scientific consensus that livestock and fertilisers are responsible for ammonia emissions \citep{Nenes2020, Thunis2021}. In Europe, around 90\% of ammonia emissions originate from the agricultural sector \citep{EEA2020}, while, in the Italian Lombardy region, shown in Figure~\ref{fig:lombardy_italy}, up to 97\% of ammonia emissions are linked to the agricultural sector \citep{INEMAR2020}. According to Lombardy Regional Environment Protection Agency \citep{Ammonia2019}, ammonia is responsible for 60\% of PM$_{10}$ concentrations in Lombardy under specific conditions.\\
\indent This paper presents an open access spatiotemporal dataset, named the Agrimonia dataset \citep{Agrimonia_dataset}, which includes several environmental variables that have been harmonised at the same spatial and temporal resolution. The dataset has been developed within the AgrImOnIA project framework (\emph{Agriculture Impact On Italian Air} - \url{https://agrimonia.net}) which aims at assessing the role of the livestock sector on the air quality in the Lombardy region. The Agrimonia dataset provides the user with a dataset with emissions data, including ammonia, agricultural information and air pollution in a common table. In general, handling spatiotemporal data from several sources is a challenge faced by many research fields and represents an interdisciplinary topic. \\ 
\indent The Agrimonia dataset could be useful for other researchers, for example, for the comparison between urban air pollution and rural air quality \citep{StrosniderEtAl2017, WenEtAl2022}. Other uses of the dataset may move toward the study of different livestock management techniques and organic products \citep{ChapmanDarby2016} or for epidemiological studies, which aim to assess the impact of agricultural emissions on the mortality attributable to air pollution~\citep{pozzer2017}. Additionally, the land use and land cover variables included in the Agrimonia dataset, provide indices of the urbanisation degree \citep{DuvernoyEtAl2018}, ecosystems conservation \citep{DeGroot2006}, and natural resources exploitation, allowing for assessing the degree of local sustainable development \citep{BastEtAl2017}. \\
\indent The rest of this paper is organised as follows: in the Methods Section, we describe the data sources and the transformations applied to harmonise the dataset, as well as the methodologies used to impute missing data and handle negative values. In the Data records Section, we describe our dataset and the various associated metadata files. Finally, the quality of the dataset and the method adopted are discussed in the Technical validation Section. 

\section*{Methods}
The Agrimonia dataset includes satellite data, model output and in situ measurements with different spatial and temporal resolutions from national and international agencies. Therefore, to combine the different datasets, a processing step is necessary. The remainder of this section describes the data sources and the harmonisation process applied to the different input data to make them homogeneous in time, with a daily resolution, and space, at the air quality station level.    

\subsection*{Source data description} 
\begin{figure}[ht]
    \centering
    \includegraphics[scale=0.6]{ 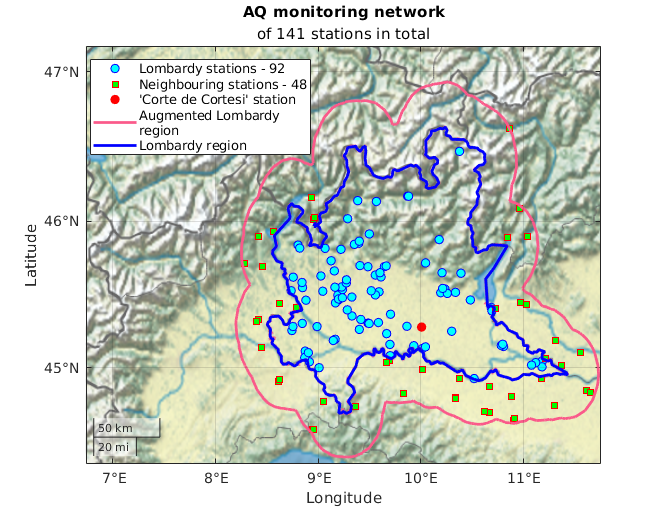}
    \caption{Administrative Lombardy region (\textit{blue} boundaries) and augmented region (\textit{pink} boundaries). A 0.3° buffer is applied to the regional boundaries to create the neighbouring area. The measurement stations of the dataset are displayed as cyan-coloured circles for the stations in Lombardy and green-coloured squares for the stations in the neighbouring area, respectively. The resulting network takes 141 stations for a total of 540 sensors. Station named `Corte de Cortesi’ (marked as a red circle) is used as a reference throughout the paper sections.  \label{aq_monitoring_network}}
\end{figure}

The data presented are related to five dimensions: air quality (AQ), weather and climate (WE), pollutants’ emissions (EM), livestock (LI) and land and soil characteristics (LA). Because geostatistical methods can use neighbouring territory information \citep{cressie2015statistics} for improving the overall predictive capability close to the borders, we take into account an area around Lombardy region by applying a 0.3° buffer over the regional borders as shown in Figure~\ref{aq_monitoring_network}. The neighbouring area intersects several regions. The various data sources used to create the Agrimonia dataset are summarised in Table~\ref{tab:data_source} and described in the following subsections that detail spatiotemporal resolution and availability. 

\begin{table}[ht] 
\newcolumntype{N}{>{\arraybackslash}X}
\begin{tabularx}{\textwidth}{m{1.6cm}m{4.5cm}m{4.5cm}NN}
\toprule
\centering\textbf{Dimension} & \centering\textbf{Source}	& \centering\textbf{Data description} & \centering\textbf{Spatial coverage} & \textbf{Temporal resolution} \\
\midrule
\multirow{4.5}{2cm}{Air quality  (AQ) } & EEA - Air pollution section & Air pollutants concentrations \citep{EEAData} & Europe & Daily, Hourly, Bi-hourly \\
\cmidrule{2-5}
& ARPA Lombardy - Air quality section& Air pollutants concentrations \citep{Maranzano2022} & Lombardy & Daily, Hourly, Bi-hourly\\
\hline
\multirow{1}{2cm}{Weather  ~(WE) } & Copernicus Climate Change Service (ERA5) & Estimates of atmospheric and land cover variables \citep{ER5Dataset,ER5LandCover} & Europe & Hourly \\[0.5cm]
\hline
\multirow{1}{2cm}{Emission  (EM)} & Copernicus Atmosphere Monitoring Service (CAMS) & Emission inventories \citep{cams} & Global & Monthly\\[0.5cm]
\hline
\multirow{1}{2cm}{Livestock ~(LI) } & National Data Bank (BDN) of the Zootechnical Registry & Livestock inventories \citep{BDN_Dataset} & Italy & Biannual\\[0.5cm]
\hline
\multirow{5.5}{2cm}{Land cover  ~(LA) } & Copernicus Land Monitoring Service (CLMS) & Corine land use classification \citep{CORINEDataset} & Europe & Only 2018 \\
\cmidrule{2-5}
& Lombardy Region Agriculture Information System (SIARL) & Cultivation type classification \citep{SIARL_Dataset} & Lombardy & Annual \\
\cmidrule{2-5}
& Fifth generation ECMWF atmospheric reanalysis (ERA5) & Estimates of land cover variables \citep{ER5LandCover} & Europe & Hourly \\
\bottomrule
\end{tabularx}
\caption{Sources of the Agrimonia dataset.\label{tab:data_source}}
\end{table}

\subsubsection*{Air quality}
The AQ data are pollutant concentrations [$\mu g/m^3 $] sampled at $S = 141$ ground-level monitoring stations, irregularly located over the augmented Lombardy region, as shown in Figure~\ref{aq_monitoring_network}. For the Lombardy area, AQ data are retrieved by the environment protection agency of the Lombardy region (ARPA Lombardy, hereinafter ARPA), while outside of Lombardy, AQ data are provided by European Environment Agency (EEA). Most of the concentrations listed in Table \ref{tab:aq_variable}, namely PM\textsubscript{2.5}, PM\textsubscript{10}, NO\textsubscript{2}, NO$_x$, CO and SO\textsubscript{2}, come from the open data system of ARPA (\url{https://www.arpalombardia.it/Pages/Aria/qualita-aria.aspx}, accessed on 5 February 2022) and are validated under EEA protocols. Instead data about NH\textsubscript{3} come from experimental monitoring campaigns, implemented by ARPA according to laboratory best practices \citep{Ammonia2019,SuperSitiLomb}, but not formally validated under the same EEA protocols. In this work, data from ARPA are collected using the ARPALData package written in R language and available on CRAN (version 1.2.3) (\url{https://cran.r-project.org/web/packages/ARPALData/index.html}, accessed on 5 February 2022). The data used for the Lombardy neighbouring areas come from the open access service of the EEA (\url{https://www.eea.europa.eu/themes/air}, accessed on 5 February 2022). To get an overview of AQ data used, Table~\ref{tab:aq_variable} summarises the pollutants selected, their sources and the number of sensors available for each pollutant.  

\begin{table}[ht]
\newcolumntype{C}{>{\centering\arraybackslash}X}
\begin{tabularx}{\textwidth}{m{1.5cm}m{6cm}m{3.7cm}CC}
\toprule
\thead{Pollutant} & \thead{Description}  &  \thead{Source}  & \textbf{Temporal resolution} & \textbf{Number of sensors}\\
\hline 
PM$_{10}$ & Particulate matter with an aerodynamic diameter
of less than 10 µm & ARPA Lombardy, EEA  & Daily, hourly & 113\\
\hline
PM$_{2.5}$ & Particulate matter with an aerodynamic diameter
of less than 2.5 µm & ARPA Lombardy, EEA  & Daily, hourly, bi-hourly & 59\\
\hline 
CO & Carbon monoxide & ARPA Lombardy, EEA  & Hourly & 60\\
\hline 
NH$_3$ & Ammonia & ARPA Lombardy & Daily, hourly & 10\\
\hline 
NO$_x$ & Nitrogen oxides & ARPA Lombardy, EEA  & Hourly & 119\\
\hline 
NO$_2$ & Nitrogen dioxide & ARPA Lombardy, EEA  & Hourly & 136\\
\hline 
SO$_2$ & Sulphur dioxide & ARPA Lombardy, EEA  & Hourly, bi-hourly & 43\\
\bottomrule
\end{tabularx}
\caption{AQ pollutants concentrations [$\mu g/m^3$] data sources, descriptions, sampling frequency and number of available sensors. \label{tab:aq_variable}}
\end{table}

At each AQ station concentration data of possibly different subsets of pollutants are gathered. For each AQ station, pollutants, spatial location, altitude, station type and other information are available in the station registry metadata file named `Metadata\_monitoring\_network\_registry.csv’ provided with the Agrimonia dataset~\citep{Agrimonia_dataset}. EEA and ARPA classify air quality stations according to the land use and emission contexts \citep{Maranzano2022}. Namely urban (U), suburban (S), rural (R) for the former and background (B), traffic (T) and industrial (I) for the latter. In particular, the $S = 141$ stations of the augmented Lombardy region are classified as 42 (UB), 3 (UI), 36 (UT), 25 (SB), 4 (SI), 1 (ST), 18 (RB) and 2 (RI).   

\subsubsection*{Weather}
Meteorological data are obtained from the Copernicus Climate Change Service (\url{https://climate.copernicus.eu/}, accessed on 27 April 2022) through the ERA5 datasets containing the numerical model output computed by the European Centre for Medium-Range Weather Forecasts (ECMWF). ERA5 is the fifth generation ECMWF reanalysis of the global climate for the past decades. The reanalysis combines model data with observations from across the world into a globally complete and consistent dataset using the laws of atmospheric science. The ERA5 datasets used here are ERA5-Single level ~\citep{ER5Dataset} and ERA5-Land~\citep{ER5LandCover}. An overview of all ERA5 sub-datasets can be found in the official ERA5 data documentation (\url{https://confluence.ecmwf.int/display/CKB/ERA5\%3A+data+documentation}, accessed on 27 April 2022). ERA5-Single level provides hourly estimates for a large number of atmospheric and land-surface quantities with a regular grid scheme at various atmosphere levels. ERA5-Land provides near-surface variables over several decades at an enhanced resolution compared to the ERA5-Single level. ERA5-Single level and ERA5-Land datasets can be downloaded through the Climate Data Store portal (\url{https://cds.climate.copernicus.eu/cdsapp#!/home}, accessed on 27 April 2022) on a regular latitude/longitude grid of 0.25° $\times$ 0.25° and 0.1° $\times$ 0.1°, respectively. To get an overview of WE variables selected from ERA5 datasets, Table~\ref{tab:we_variables} summarises the WE variables, their sources, descriptions and units.\\
\indent Relative humidity is useful for studying air quality, so we complete the weather variables by calculating relative humidity. Using the temperature ($T$) and dew point temperature ($T_{dew}$), we compute the relative humidity ($RH$) using the August-Roche-Magnus approximation formula \citep{alduchov1996improved}:

\begin{equation}
    RH = 100 \times \exp\left(\frac{17.625 \times T_{dew}}{243.04+T_{dew}} - \frac{17.625 \times T}{243.04+T}\right)
\end{equation}

\begin{table}[ht] 
\newcolumntype{C}{>{\centering\arraybackslash}X}
\begin{tabularx}{\textwidth}{m{2cm} m{3.8cm} m{7.6cm} C}
\toprule
\thead{Dataset} & \thead{Variable}  & \thead{Description} & \textbf{Unit}\\
\hline 
\multirow{12}{2cm}{ERA5 Land}
& 10 m u-component of wind & Eastward component of the wind at 10 metres altitude & $m s^{1}$  \\
\cmidrule{2-4}
& 10 m v-component of wind & Northward component of the wind at 10 metres altitude & $m/s^{1}$ \\
\cmidrule{2-4}
& 2 m dewpoint temperature  & Temperature to which the air, at 2 metres above the surface of the Earth, would have to be cooled for saturation to occur &  $K$  \\
\cmidrule{2-4}
& 2 m temperature  & Temperature of air at 2 metres above the surface of land, sea or inland waters &  $K$  \\
\cmidrule{2-4}
& Total precipitation & Accumulated liquid and frozen water, comprising rain and snow that falls to the Earth's surface & $m$ \\
\cmidrule{2-4}
& Surface net solar radiation & Amount of solar radiation that reaches a horizontal plane at the surface minus the amount reflected by the Earth’s surface & $J/m^{2}$  \\
\midrule 
\multirow{9}{2cm}{ERA5 Single Level} 
& 100 m u-component of wind & Eastward component of the wind at 100 metres altitude & $m/s^{1}$ \\
\cmidrule{2-4}
& 100 m v-component of wind &  Northward component of the wind at 100 metres altitude & $m/s^{1}$\\
\cmidrule{2-4}
& Boundary layer height &  Depth of air next to the Earth’s surface that is most affected by the resistance to the transfer of momentum, heat or moisture across the surface & $m$ \\
\cmidrule{2-4}
& Surface pressure & Pressure (force per unit area) of the atmosphere at the surface of land  & $Pa$ \\
\cmidrule{2-4}
& Precipitation type & Type of precipitation on the Earth's surface. Values of precipitation type are: no precipitation (0), rain (1), freezing rain (3), snow (5), wet snow (6), mixture of rain and snow (7), ice pellets (8) & $Categorical$ \\
\bottomrule
\end{tabularx}
\caption{WE variables selected from ERA5 datasets.\label{tab:we_variables}}
\end{table}

\subsubsection*{Emissions}
The Copernicus Atmosphere Monitoring Service (CAMS) implemented by the ECMWF is one of the most recent global databases covering anthropogenic source emissions. CAMS datasets are compiled emission inventories for many atmospheric compounds developed for the years 2000-2022 \citep{granier2019copernicus,cams}. These inventories are based on a combination of existing datasets and new information, describing anthropogenic emissions from fossil fuel use on land, natural emissions from vegetation, soil and more. The anthropogenic emissions on land are further separated into specific activity sectors (e.g. traffic, agriculture). Pollutant emissions data are provided by the CAMS-anthropogenic emissions dataset (\url{https://permalink.aeris-data.fr/CAMS-GLOB-ANT}, accessed on 27 April 2022), which contains monthly global anthropogenic and natural emissions from 36 sources on a regular grid level. The anthropogenic sources are divided into 20 sectors (including agriculture and livestock) with a spatial resolution of 0.1° $\times$ 0.1°. Table~\ref{tab:em_variables} summarises the emission variables selected, their origins, descriptions and units.\\ 

\begin{table}[ht] 
\newcolumntype{C}{>{\arraybackslash}X}
\begin{tabularx}{\textwidth}{m{1cm} m{3.3cm} m{9cm} C}
\toprule
\thead{Variable} & \thead{Emission origin} & \thead{Description} & \textbf{Unit} \\
\midrule 
\multirow{5}{2cm}{NH$_3$} & Livestock sector & Emissions of NH\textsubscript{3} originating from the livestock sector for manure management  &   $kg / (m^2s)$  \\
\cmidrule{2-4}
& Agriculture soils & Emissions of NH\textsubscript{3} originating from agriculture soils    &   $kg/(m^2s)$ \\
\cmidrule{2-4}
& Waste burning & Emissions of NH\textsubscript{3} originating from agriculture waste burning    &   $kg/(m^2s)$   \\
\cmidrule{2-4}
& Total & Total emissions of NH\textsubscript{3} across all sectors     &   $kg/(m^2s)$               \\
\hline
\multirow{2}{2cm}{NO$_x$} & Road transportation & Emissions of NO\textsubscript{x} from on road transportation   &     $kg/(m^2s)$  \\
\cmidrule{2-4}
& Total & Emissions of NO\textsubscript{x} across all sectors   &     $kg/(m^2s)$              \\
\hline
\multirow{1}{2cm}{SO$_2$} & Total & Emissions of SO\textsubscript{2} across all sectors   &     $kg/(m^2s)$   \\
\bottomrule
\end{tabularx}
\caption{EM variables selected from the CAMS-anthropogenic emissions dataset.\label{tab:em_variables}}
\end{table}

\subsubsection*{Livestock}
\begin{figure}[ht]
\centering
      \begin{subfigure}[c]{.475\linewidth}
        \centering         
        \includegraphics[width=\linewidth]{ 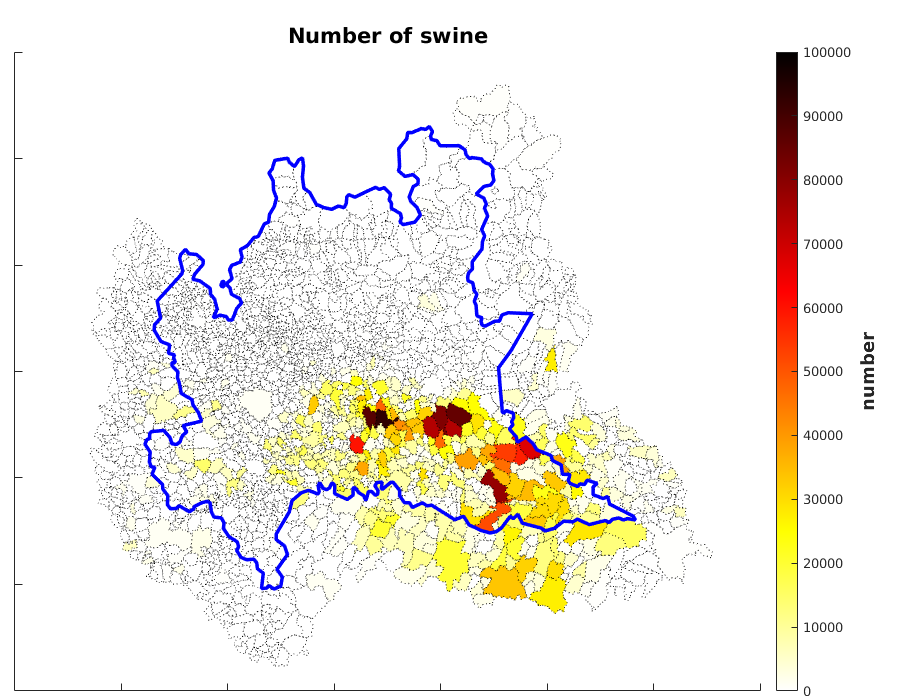}
         \caption{ \centering \label{fig:bdn_pigs_abs}}
    \end{subfigure}
     \begin{subfigure}[c]{.475\linewidth}
         \includegraphics[width=\linewidth]{ 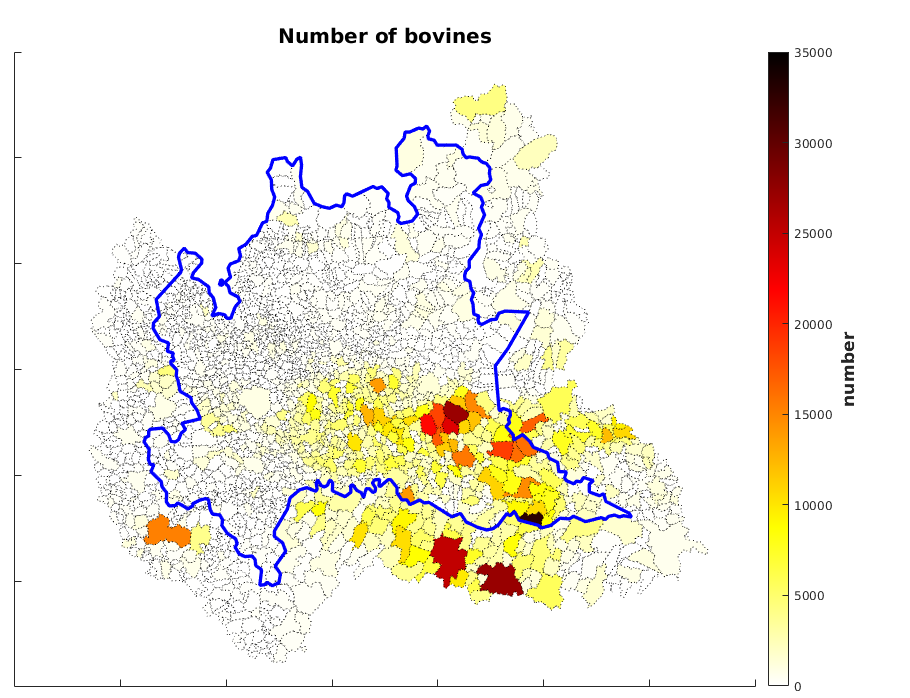}
         \caption{ \centering \label{fig:bdn_bovine_abs}}
    \end{subfigure}
\caption{Number of swine \textbf{(a)} and bovines \textbf{(b)} in the augmented Lombardy region and neighbouring area, aggregated at the municipal level on 31 December 2021. \label{fig:li_maps}}
\end{figure}

Slurry and manure have the highest concentration of nitrogen ($N$) in a soluble form, which is a precursor of ammonia gases. The animal categories that are the most responsible for $N$ emissions are swine and bovines\citep{ministero}, for this reason only the data related to these species are included in the Agrimonia dataset. Information about livestock is obtained by the Italian National Data Bank of the Zootechnical Registry (BDN) \citep{BDN_Dataset}. The BDN dataset is derived from the livestock census and includes information on the animal population of zootechnical interest present in Italy, its distribution in the territory and its characteristics. The BDN dataset is managed by the Directorate General for Animal Health and Veterinary Medicines (\url{https://www.salute.gov.it/portale/temi/p2_5.jsp?lingua=italiano&area=sanitaAnimale&menu=tracciabilita}, accessed on 15 February 2022), translated as \textit{Direzione Generale della Sanità Animale e dei Farmaci Veterinari} of the Italian Ministry of Health and represents the official source of data on livestock, both for the control authorities and users. The BDN dataset is accessible through the “statistics” section of the BDN portal (\url{https://www.vetinfo.it/j6_statistiche/index.html#/}, accessed on 15 February 2022). The BDN data are updated every six months and aggregated at the municipality level. From BDN, we take the municipal number of swine and bovines for the augmented Lombardy region. Figure~\ref{fig:li_maps} shows the number of swine and bovines in the augmented Lombardy region, which are both particularly high in the Southeastern areas. 

\subsubsection*{Land}
Land cover, land and soil use are considered important factors to assess the agriculture impact on PM \citep{Nenes2020, SPASTA2021}. The data about land are retrieved from different sources: for land cover variables (only high and low vegetation index) we use the ERA5-Land \citep{ER5LandCover} dataset already introduced in the Weather Section. The land use variables are provided by the Corine Land Cover (CLC) \citep{CORINEDataset} dataset while the soil use variables are given by the Lombardy Region Agriculture Information System (SIARL)\citep{SIARL_Dataset} dataset. The CLC dataset, handled by Copernicus Land Monitoring Service (CLMS) (\url{https://www.copernicus.eu/en/copernicus-services/land}, accessed on 27 April 2022), is available only for the year 2018 and consists of an inventory of land use in 44 classes within a minimum mapping unit of 25 hectares. The classes are organised in the hierarchical 3-level CLC nomenclature \citep{CORINE_manual} based on the classification of satellite images. The CLC dataset can be downloaded from the CLMS portal (\url{https://land.copernicus.eu/pan-european/corine-land-cover}, accessed on 27 April 2022). Along with land cover and land use, we additionally include soil use and cultivation type information provided by the SIARL dataset, which is updated until 2019. Table \ref{tab:la_variable} summarises the land cover, land and soil use variables selected from the ERA5-Land, CLC and SIARL datasets, respectively. In the metadata files provided with the Agrimonia dataset \citep{Agrimonia_dataset}, the class labels for CLC and SIARL datasets are available in the files named `Metadata\_LA\_CORINE\_labels.csv’, for the CLC classes, and `Metadata\_LA\_SIARL\_labels.csv’ for the SIARL classes. 

\begin{table}[ht] 
\newcolumntype{C}{>{\centering\arraybackslash}X}
\begin{tabularx}{\textwidth}{m{2cm} m{3.5cm} m{8cm} C}
\toprule
\thead{Dataset} & \thead{Variable} & \thead{Description} & \textbf{Unit}\\
\midrule
\multirow{3}{2.3cm}{ERA5-Land (land cover)}
& Low vegetation index & One-half of the total green leaf area per unit horizontal ground surface area for low vegetation type & $m^2/m^{2}$ \\
\cmidrule{2-4}
& High vegetation index & One-half of the total green leaf area per unit horizontal ground surface area for high vegetation type & $m^2/m^{2}$ \\
\midrule
CLC ~~~~~~~~~~(land use) & Third-level of land use  & Classification of land use with 44 classes & \textit{Categorical}  \\ 
\midrule
SIARL ~~~~~~(soil use) & Soil cultivation type & Classification of soil use with 21 classes & \textit{Categorical} \\ 
\bottomrule
\end{tabularx}
\caption{LA variables selected from the ERA5, CLC and SIARL datasets.\label{tab:la_variable}}
\end{table}

\subsection*{Data harmonisation and processing}

Since the previous section discussed the input data’s different spatial and temporal resolutions, we introduce here the methods used to harmonise the data before merging them into the Agrimonia dataset. We also consider missing data imputation and some variable transformations. In the metadata file named `Metadata\_Agrimonia.csv’, provided along with the Agrimonia dataset \citep{Agrimonia_dataset}, columns 9 - 14 summarise the original spatial and temporal resolutions and the transformations converting the variables to the same spatial and temporal resolution, i.e., daily quantities at the station level.

\subsubsection*{Air quality}
The AQ data, from ARPA and EEA networks, have some issues related to missing measurements; that is, they have both short and prolonged periods with stations turned off for maintenance, instrument calibration or other reasons. Furthermore, measurements could be taken at irregular intervals, or a sampling policy change could take place. Measurements not validated by the environmental agencies and negative values are considered as missing values (`NaN’).\\
\indent As reported in Table~\ref{tab:aq_variable}, both daily, bi-hourly and hourly AQ measurements are present in the network. Since some time series are hybrid (i.e. they have different time resolutions) we have considered the distribution of the time gaps between the measures. A time series with constant temporal resolution (e.g. daily) has a unimodal gap frequency distribution. Vice versa, a time series with a hybrid time resolution results in a bimodal gap frequency distribution. \\
\indent Since the presence of several missing values in a day could introduce a bias in the daily average, we implemented the following algorithm. For each hourly and bi-hourly time series, missing values are imputed using a state-space model~\citep{ssmodel} and the relative Kalman smoother~\citep{Kalman} which provides an estimate of the missing data and their uncertainty. Next, hourly and bi-hourly time series are averaged over each day. Days with a gap larger than six hours are set to missing. The Kalman smoother uncertainty associated with the hourly estimate is propagated to the daily average, thus providing daily uncertainties due to the missing data imputation. Details on this approach are discussed in the Technical validation Section. The resulting imputation uncertainty is reported in the metadata file `Metadata\_AQ\_imputation\_uncertainty.csv’ provided along with the Agrimonia dataset \citep{Agrimonia_dataset}. Table~\ref{tab:Agrimonia_aq_variable} summarises AQ variables in the Agrimonia dataset providing name, spatial and temporal transformations and time resolution.

\begin{table}[ht]
\newcolumntype{C}{>{\centering\arraybackslash}X}
\begin{tabularx}{\textwidth}{m{2.5cm} m{8cm} C}
\toprule
\thead{Variable name}   &  \thead{Description}  & \textbf{Temporal transformation} \\
\midrule
 AQ\_pm\_10 & Particulate matter with an aerodynamic diameter of less than 10 µm  & Kalman smoother and daily mean  \\
\midrule
 AQ\_pm\_2.5 & Particulate matter with an aerodynamic diameter of less than 2.5 µm  & Kalman smoother and daily mean  \\
\midrule
 AQ\_co & Carbon monoxide  & Kalman smoother and daily mean  \\
\midrule
 AQ\_nh3 & Ammonia  & Kalman smoother and daily mean  \\
\midrule
 AQ\_nox & Several oxides of nitrogen  & Kalman smoother and daily mean  \\
\midrule 
AQ\_no2 & Nitrogen dioxide  & Kalman smoother and daily mean  \\
\midrule
 AQ\_so2 & Sulphur dioxide  & Kalman smoother and daily mean  \\
\bottomrule
\end{tabularx}
\caption{AQ pollutants concentrations [$\mu g/m^3 $] in the Agrimonia dataset. All AQ variables included in the Agrimonia dataset are harmonised to the daily time resolution. More details on the transformation process can be found in the metadata file named named `Metadata\_Agrimonia.csv’ available with Agrimonia dataset~\citep{Agrimonia_dataset}. \label{tab:Agrimonia_aq_variable}}
\end{table}

\subsubsection*{Weather}
\begin{figure}[ht]
\centering
      \begin{subfigure}[c]{.475\linewidth}
        \centering
         \includegraphics[width=\linewidth]{ 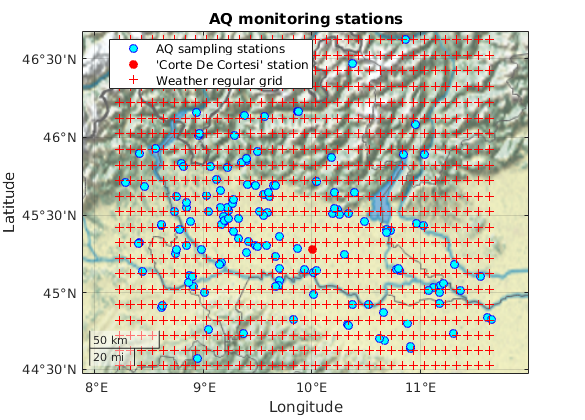}
         \caption{ \centering\label{fig:we_grid}}
     \end{subfigure}
     \begin{subfigure}[c]{.475\linewidth}
         \centering
         \includegraphics[width=\linewidth]{ 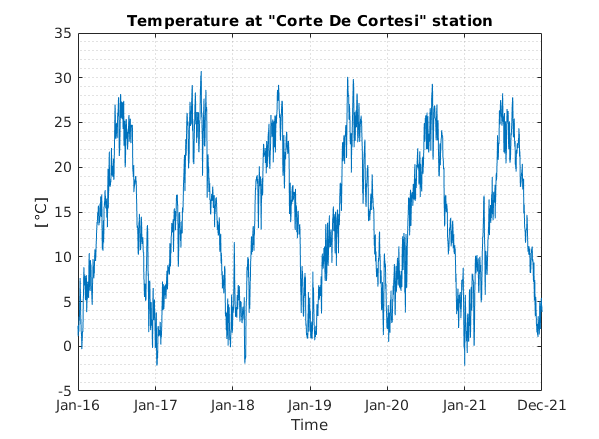}
         \caption{\centering \label{fig:we_temp_2m}}
     \end{subfigure}
\caption{\textbf{(a)}: The irregularly located AQ monitoring stations  (cyan circles) and the weather grid centres (red `+’ symbols). \textbf{(b)}: Daily 2 m temperature time series (WE\_temp\_2m) from 2016 to 2021 for the monitoring station named `Corte De Cortesi’ (see red point in Figure \ref{fig:we_grid}).}
\end{figure}

This section describes in detail the harmonisation process for the WE variables (see Table~\ref{tab:Agrimonia_we_variable}).
The data about WE come from ERA5 datasets and are given by hourly reanalysis estimates in a regular grid format. It should be noted that the ERA5 value refers to the grid cell average, while the coordinates refer to the centre of the cell. Because the data stem from a model, there is no problem with missing and negative values. Some variables need to be preprocessed to be more informative. We summarise the two preprocessing steps for the original variables, as follows: the wind speed is calculated as the Euclidean norm of the wind vector with \textit{u-} and \textit{v-} components; the wind direction is discretised using the classical 8-wind rose: North (N), North-east (NE), East (E), South-east (SE), South (S), South-west (SW), West (W), North-west (NW); the temperature is converted from Kelvin to Celsius degrees. \\
\indent The transformation of weather variables to create daily time series is composed of two different stages. The first one is to create an hourly weather time series related to each AQ monitoring station, while the second consists of computing daily time series from hourly time series. The first step is necessary because the AQ station is misaligned concerning weather data, as shown in Figure~\ref{fig:we_grid}. To associate weather time series to each AQ station, we use the \emph{inverse distance weighted} (IDW) interpolation algorithm  \citep{shepard1968two}. The IDW algorithm is based on the Euclidean distance between the localisation of the stations and grid cells’ centres. For each station, we consider the four nearest grid cells. The IDW power parameter, which controls the weight of the cell values on the interpolated values based on their distance from the localisation of the station, is set to one. After the hourly time series is created using the IDW approach for each station, we convert the temporal resolution from hourly to daily using different ensemble functions according to the variable type \citep{cameletti2011comparing}. Table~\ref{tab:Agrimonia_we_variable} lists the weather variables in the Agrimonia dataset while Figure~\ref{fig:we_temp_2m} shows an example of time series obtained using the IDW approach.  

\begin{table}[ht]
\newcolumntype{C}{>{\centering\arraybackslash}X}
\begin{tabularx}{\textwidth}{m{4.6cm}m{7.4cm}m{1.8cm}C}
\toprule
\thead{Variable name}   &  \thead{Description} & \centering\textbf{Aggregation function}  & \textbf{Units}\\
\midrule
WE\_temp\_2m &  Temperature of air at 2 m above the surface of land, sea or inland water  & Daily mean  &  °$C$  \\
\midrule
WE\_wind\_speed\_10\_mean & Mean intensity of the wind speed at a height of 10 m above the surface of the Earth   & Daily mean & $m/s$ \\
\midrule
WE\_wind\_speed\_10\_max & Max intensity of the wind speed at a height of 10 m above the surface of the Earth & Daily max &  $m/s$ \\
\midrule
WE\_mode\_wind\_direction\_10m & Direction of the wind intensity at a height of 10 m above the surface of the Earth  & Daily mode &  \textit{Categorical} \\ 
\midrule
WE\_tot\_precipitation & The accumulated liquid and frozen water, comprising rain and snow, that falls to the Earth’s surface    & Daily sum  &  $m$  \\
\midrule
WE\_precipitation\_t & The type of precipitation on the Earth's surface, at the specified time & Daily mode  &   \textit{Categorical}  \\
\midrule
WE\_surface\_pressure & The pressure (force per unit area) of the atmosphere at the surface of land, sea and inland water    & Daily mean  &  $Pa$  \\
\midrule
WE\_solar\_radiation & Amount of solar radiation that reaches a horizontal plane at the surface minus the amount reflected by the Earth’s surface    & Daily max  &  $ J/m^2$ \\
\midrule
WE\_wind\_speed\_100\_mean & Mean intensity of the wind speed at a height of 10 m above the surface of the Earth & Daily mean  &  $m/s$  \\
\midrule
WE\_wind\_speed\_100\_max & Max intensity of the wind speed at a height of 10 m above the surface of the Earth & Daily max  &  $m/s$  \\
\midrule
WE\_mode\_wind\_direction\_100m & Direction of the wind intensity at a height of 100 m above the surface of the Earth & Daily mode  &   \textit{Categorical} \\
\midrule
WE\_blh\_layer\_max & The maximum depth of air next to the Earth’s surface that is the most affected by the resistance to the transfer of momentum, heat or moisture across the surface    & Daily max  &  $m$  \\
\midrule
WE\_blh\_layer\_min & The minimum depth of air next to the Earth’s surface that is the most affected by the resistance to the transfer of momentum, heat or moisture across the surface    & Daily min  &  $m$  \\
\midrule
WE\_rh\_min &  Maximum amount of water vapour present in air expressed as a percentage of the amount needed for saturation at the same temperature    & Daily min  &   \% \\
\midrule
WE\_rh\_mean &  Mean amount of water vapour present in air expressed as a percentage of the amount needed for saturation at the same temperature    & Daily mean  &   \%  \\
\midrule
WE\_rh\_max & Minimum amount of water vapour present in air expressed as a percentage of the amount needed for saturation at the same temperature    & Daily max  &   \%  \\
\bottomrule
\end{tabularx}
\caption{WE variables included in the Agrimonia dataset. More detail on the transformation process can be found in the metadata file named `Metadata\_Agrimonia.csv’ available with Agrimonia dataset~\citep{Agrimonia_dataset}. \label{tab:Agrimonia_we_variable}}
\end{table}

\clearpage

\subsubsection*{Emissions}
This section describes the harmonisation process for the EM variables summarised in Table~\ref{tab:Agrimonia_em_variable}. The data about EM are from the CAMS datasets with a monthly temporal resolution and on a regular grid. As done for the WE variables, we performed a two-step transformation process to create daily emission time series. In the first step, we use the same IDW approach described for WE variables to compute emission values related to each monitoring station with monthly resolution. In the second step, we use spline interpolation techniques to convert the series to the same daily temporal resolution. To avoid oscillations, overshoots, edge effects and negative values, we use \textit{piecewise cubic Hermite interpolating polynomials} (PCHIP) \citep{pchip}. As discussed in more detail in the Technical validation Section, this method interpolates the data smoothly, while retaining the data’s shape and monotonicity. \\

\begin{table}[ht]
\begin{tabularx}{\textwidth}{m{4.5cm}m{10cm}}
\toprule
\thead{Variable name}   &  \thead{Description} \\
\midrule
EM\_nh3\_livestock\_mm & Emissions of NH$_3$ originating from the livestock sector for the manure management  \\
\midrule
EM\_nh3\_agr\_soils & Emissions of NH$_3$ originating from agriculture soils   \\
\midrule
EM\_nh3\_agr\_waste\_burn & Emissions of NH$_3$ originating from the burning of agriculture waste \\ 
\midrule
EM\_nh3\_sum & Total emissions of NH$_3$ across all sectors (anthropogenic) \\
\midrule
EM\_nox\_traffic & Emissions of NO$_x$ from the on-road transportation  \\
\midrule
EM\_nox\_sum & Emissions of NO$_x$ across all sectors (anthropogenic) \\
\midrule
EM\_no2\_sum & Total emissions of SO$_2$ across all sectors (anthropogenic) \\
\bottomrule
\end{tabularx}
\caption{Pollutant EM variables [$mg/m^3$] present in the Agrimonia dataset with daily temporal resolution. For each EM variable we use IDW function as spatial transformation and PCHIP interpolation function to transform form monthly to daily temporal resolution. More detail on the transformation process can be found in the metadata file named `Metadata\_Agrimonia.csv’ available with Agrimonia dataset~\citep{Agrimonia_dataset}. \label{tab:Agrimonia_em_variable}}
\end{table}

\subsubsection*{Livestock}
\begin{figure}[ht]
\centering
      \begin{subfigure}[c]{.475\linewidth}
        \centering         
        \includegraphics[width=\linewidth]{ 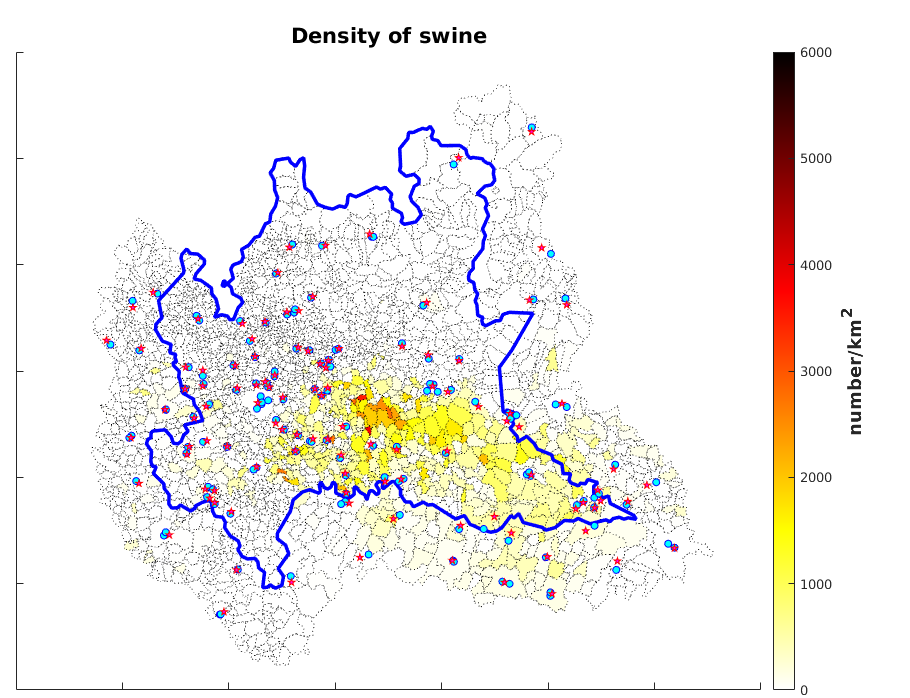}
         \caption{ \centering \label{fig:bdn_pigs_den}}
    \end{subfigure}
     \begin{subfigure}[c]{.475\linewidth}
         \includegraphics[width=\linewidth]{ 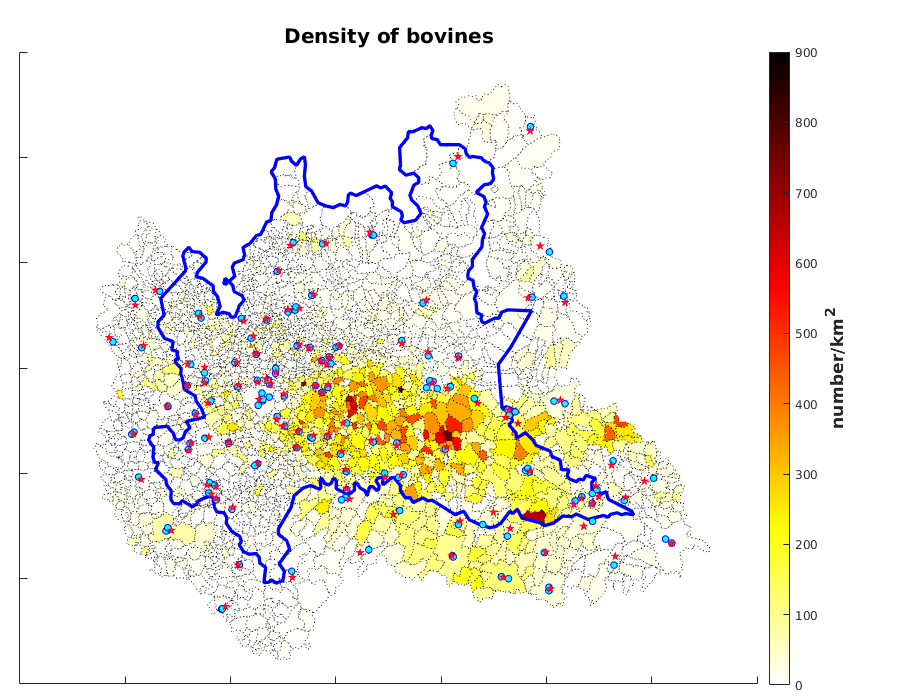}
         \caption{ \centering \label{fig:bdn_bovine_den}}
    \end{subfigure}
\caption{Swine \textbf{(a)} and bovines \textbf{(b)} density over the augmented Lombardy region on 31 December 2021. The AQ stations (cyan circle) are spread randomly over the studied area. Each station is associated with the the municipality centroid to which it belongs (red start). Based on this, stations in the same municipality share the same municipality centroid so they have the same livestock time series.\label{fig:bdn_aq}}
\end{figure}

In this section, we describe the harmonisation process for the LI variables summarised in Table~\ref{tab:Agrimonia_li_variable}. The data related to the livestock sector are retrieved from the BDN dataset, which provides the number of bovines and swine aggregated at the municipality level. The BDN dataset is updated every six months, in June and in December. As a result, for each municipality, a time series of 12 values are available. Each AQ station is associated with the time series of the municipality to which it belongs, see Figures \ref{fig:bdn_pigs_den} and \ref{fig:bdn_bovine_den} for swine and bovines, respectively. Due to the particular municipality shape, the station named `Vallelaghi\_T1191A' is within a municipality whose centroid is outside of the considered augmented domain, therefore the value of the closest municipality in the area considered is taken. \\
\indent As done for the CAMS data, the PCHIP interpolation is used to increase the temporal resolution from biannual to daily (for more details, see the Technical validation Section). Once the interpolation function has been chosen, it is possible to evaluate it over the entire time horizon, particularly for all daily instants. To reduce edge effects, we use the value for December 31, 2015, as the first starting value for the time series. Subsequently, the municipal animal density is calculated by dividing the animal count by the area of the station municipality (expressed in $km^2$). In this way, we obtain the daily time series of swine and bovines density for each monitoring station. See Table~\ref{tab:Agrimonia_li_variable} for a summary of the livestock variables in the Agrimonia dataset.

\begin{table}[ht]
\begin{tabularx}{\textwidth}{m{4.5cm}m{10cm}}
			\toprule
			\thead{Variable name}   &  \thead{Description} \\
			\midrule
LI\_pigs & Municipal density of swine related to AQ stations \\
\midrule
LI\_bovine & Municipal density of bovines related to AQ stations \\
\bottomrule
\end{tabularx}
\caption{LI variables in the Agrimonia dataset with daily temporal resolution. Information on the number of swine and bovines is expressed as a density with respect to the municipal area: $number/km^2$. More detail on the transformation process can be found in the metadata file named `Metadata\_Agrimonia.csv’ available with Agrimonia dataset~\citep{Agrimonia_dataset}. \label{tab:Agrimonia_li_variable}}
\end{table}

\subsubsection*{Land}
This section describes in detail the harmonisation process for the LA variables, which are summarised in Table~\ref{tab:Agrimonia_la_variable}. The data related to land cover and land and soil use are given by the ERA5-Land, CLC and SIARL datasets, respectively, describing the land and soil over time. Considering that high and low vegetation indices from ERA5-Land have a daily resolution over a spatial regular grid, we use the same IDW approach described for WE variables to create the daily time series associated with each AQ station. Information about land use is relatively stable over time. For the CLC dataset, we take the 2018 data and keep them constant for the period from 2016 to 2021. For the SIARL dataset, the values are annual until 2019 (for more detail on land use and land cover, see the Technical validation Section).
The CLC provides categorical data in polygons while SIARL does so on a regular grid. In both cases, each AQ station is associated with the polygon or the cell to which it belongs. In the metadata files provided along with the Agrimonia dataset \citep{Agrimonia_dataset}, the class labels for CLC and SIARL datasets are available in the files named `Metadata\_LA\_CORINE\_labels.csv’ and `Metadata\_LA\_SIARL\_labels.csv’, respectively. In this way, we obtain daily piecewise constant functions for land cover and soil use associated with each AQ station.

\begin{table}[ht]
\newcolumntype{C}{>{\centering\arraybackslash}X}
\begin{tabularx}{\textwidth}{m{2cm}m{7cm}CCC}
\toprule
\thead{Variable name}   &  \thead{Description} & \centering\textbf{Spatial transformation} & \centering\textbf{Temporal transformation} & \textbf{Unit} \\
\midrule
LA\_hvi & One-half of the total green leaf area per unit horizontal ground surface area for high vegetation type & IDW & None &  $ m^2/m^2 $ \\
\midrule
LA\_lvi & One-half of the total green leaf area per unit horizontal ground surface area for low vegetation type & IDW & None  &  $ m^2/m^2 $  \\
\midrule
LA\_land\_use & CORINE Land Cover - Land use across 44 sectors & None & None & \textit{Categorical}  \\
\midrule
LA\_soil\_use & SIARL Lombardy - Lombardy soil use across 21 sectors & None & None & \textit{Categorical}  \\
\bottomrule
\end{tabularx}
\caption{LA variables in Agrimonia dataset with daily time resolution. More detail on the transformation process and label for categorical variables can be found in the metadata files named `Metadata\_Agrimonia.csv’, `Metadata\_LA\_CORINE\_labels.csv’ and `Metadata\_LA\_SIARL\_labels.csv’, respectively, provided with Agrimonia dataset~\citep{Agrimonia_dataset}. \label{tab:Agrimonia_la_variable}}
\end{table}

\section*{Data records}

The output dataset has been built by joining the daily time series related to the air quality (AQ), weather (WE), emission (EM), livestock (LI) and land (LA) variables discussed in the previous sections and referred to the same AQ monitoring station for the Lombardy region augmented by the 0.3° buffer, depicted in Figure \ref{fig:lombardy_italy}. The dataset and the metadata files are available on the Zenodo repository \citep{Agrimonia_dataset} as follows: 

\begin{itemize}
    \item \textit{Agrimonia\_Dataset.csv}: this is the Agrimonia output dataset joining the daily time series, at station locations, related to the AQ (see Table \ref{tab:Agrimonia_aq_variable}), WE (see Table \ref{tab:Agrimonia_we_variable}), EM (see Table \ref{tab:Agrimonia_em_variable}), LI (see Table \ref{tab:Agrimonia_li_variable}) and LA (see Table \ref{tab:Agrimonia_la_variable}) variables. In order to simplify the access to variables in the Agrimonia dataset, the variable name starts with the dimension of the variable, e.g., the name of the variables related to the AQ dimension starts with `AQ\_’. Missing data are denoted by the `NaN’ value. The Agrimonia dataset has $S = 141$ monitoring stations and $T = 2192$ days between 1st January 2016 and 31st December 2021. The dataset is characterised by: $S \times T$ rows, each of which is uniquely identified by the pair station code and date in ``YYYY-MM-DD'' format. It contains 41 columns including the following block:  header (stations’ code, latitude and longitude, date and altitude), AQ (7 columns), WE (16 columns), EM (7 columns), LI (2 columns) and LA (4 columns). This file is made available also in the \textit{.mat} and \textit{.Rdata} format for MATLAB and R software, respectively.
    
    \item \textit{Metadata\_Agrimonia.csv}: this is the main Agrimonia metadata file and provides further information for the sources used, variables imported, transformations applied and Agrimonia variables. 
    
    \item \textit{Metadata\_monitoring\_network\_registry.csv}: it contains details about the AQ monitoring stations including station type, NUTS3 code, environment type, altitude, monitored pollutants and others. Each row represents a single sensor.
    
    \item \textit{Metadata\_AQ\_imputation\_uncertainty.csv}: it contains the estimate of the daily uncertainty due to missing data imputation for the AQ time series. In particular, for each AQ variable, days without missing hours have zero uncertainty, days with one or more imputed hours have a number resulting from the propagation of the uncertainty in the daily averaging, and days with a `NaN’ in the concentrations have a `NaN’ also in the uncertainty. 
    
    \item \textit{Metadata\_LA\_CORINE\_labels.csv}: it contains labels and descriptions associated with the CLC land variables (column CORINE code).   
    
    \item \textit{Metadata\_LA\_SIARL\_labels.csv}: it contains labels and descriptions associated with the SIARL land variables (column SIARL code).
\end{itemize}

The Agrimonia dataset is consistent with the following reference systems: World Geodetic System 1984 (WGS84) \citep{wgs84} for geo-referentiation, Coordinated Universal Time (UTC) for time referentiation and the International System of Units (SI) for the metric system, except for the temperature expressed in Celsius degrees ($^\circ C$). The values in the dataset are represented in scientific notation with 4 significant digits here considered sufficient for statistical modelling purposes. The coordinates (latitude, longitude) have a fixed point representation with 9 significant digits for the correct identification of the stations’ locations.

\section*{Technical validation}

\subsection*{Validation of AQ imputation methods}
\begin{figure}[ht]
\centering
      \begin{subfigure}[c]{.483\linewidth}
        \centering         
        \includegraphics[width=\linewidth]{ 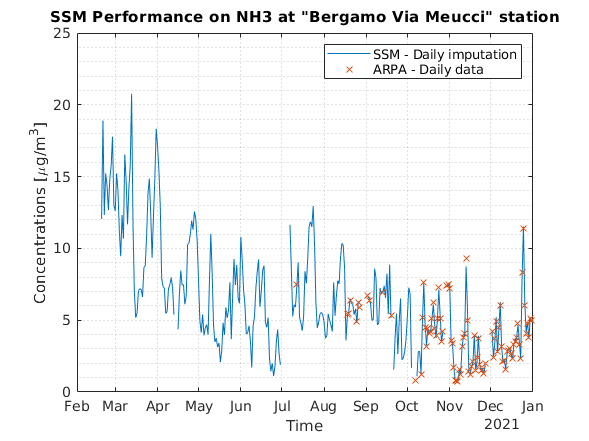}
         \caption{ \centering \label{fig:aq_nh3}}
    \end{subfigure}
     \begin{subfigure}[c]{.483\linewidth}
         \includegraphics[width=\linewidth]{ 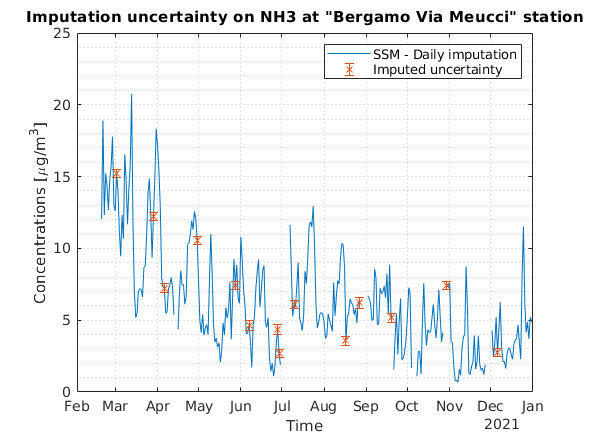}
         \caption{ \centering \label{fig:aq_nh3_unc}}
    \end{subfigure}
\caption{ \textbf{(a)}: Imputed and raw daily data for the monitoring station named `Bergamo Via Meucci’ (RMSE = $0.1710$). \textbf{(b)}: Imputed daily data with associated imputed uncertainty ($\pm 2\sigma_d$) for the monitoring station named `Bergamo Via Meucci’. \label{fig:aq_nh3_val}}
\end{figure}

The missing values imputation for the hourly and bi-hourly time series is performed using the State-Space Model (SSM)~\citep{ssmodel} and the relative Kalman smoother~\citep{Kalman}. For any hour $t \in \{1,\cdots,24 T$\}, where $T$ is the number of days as before, let $x_t$ be the scalar state describing the dynamics of the underlying AQ ``true'' concentrations and let $y_t$ be the scalar observation of the observed hourly AQ series. Moreover, let $u_t$ and $\epsilon_t$ be Gaussian white noises with unit-variance representing the innovation and measurement error, respectively, with $u_t$ and $\epsilon_t$ uncorrelated. The SSM here used for missing imputations is defined by:  

\begin{equation}
\label{eq:ssm}
    \begin{cases}
    x_t = \alpha x_{t-1} + bu_t \\
    y_t = x_t + \epsilon_t 
    \end{cases}
\end{equation}

\noindent where the $\alpha$ and $b$ parameters describe the dynamics and the additive error structure on the state $x_t$, respectively. Both $\alpha$ and $b$ are estimated for each hourly time series using numerical optimisation of the likelihood function with initial values set to one. Assuming the hourly state errors $x_t - \hat{x_t}$ uncorrelated, we propagate the imputed uncertainty given by the smoother, through the mean of the generic day ($d$) as $\sigma_d = \sqrt{ 1/24^2 \sum_{t \in d} Var(x_t\mid y_1,...,y_{24 T})}$ where $Var(x_t\mid y_1,...,y_{24 T})$ is the variance of the smoothed states $x_t$ for the hour $t$ considered during the daily mean. \\
\indent A validation experiment concerns the station called `Bergamo Via Meucci’. ARPA Lombardy provides the data at both hourly and daily resolution for this station. The hourly time series is longer but has several missing items. So, we verify the performance of the missing imputation process by comparing the daily data obtained through the Kalman smoother and the daily data provided by the agency. Figure~\ref{fig:aq_nh3} shows the two daily time series: the blue line depicts our method and the orange crosses are the ARPA Lombardy data. Figure~\ref{fig:aq_nh3_unc} shows the imputed uncertainty of daily average concentrations computed from imputed hourly time series. It can be observed that the time series obtained by our missing imputation method is very close to the daily one, with the Root Mean Square Error (RMSE) equal to $0.1710$.

\begin{figure}[ht]
\centering
      \begin{subfigure}[c]{.483\linewidth}
        \centering         
        \includegraphics[width=\linewidth]{ 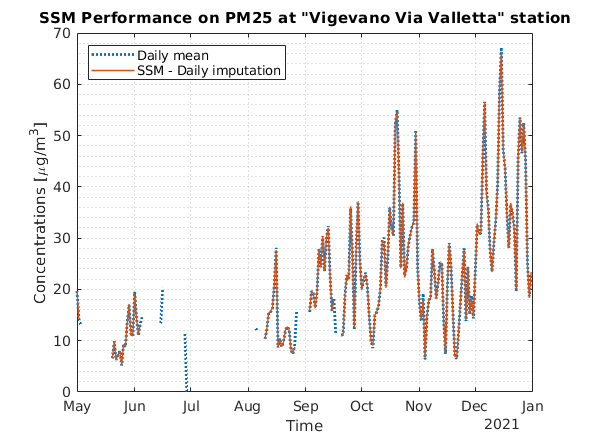}
         \caption{ \centering \label{fig:aq_pm25}}
    \end{subfigure}
     \begin{subfigure}[c]{.483\linewidth}
         \includegraphics[width=\linewidth]{ 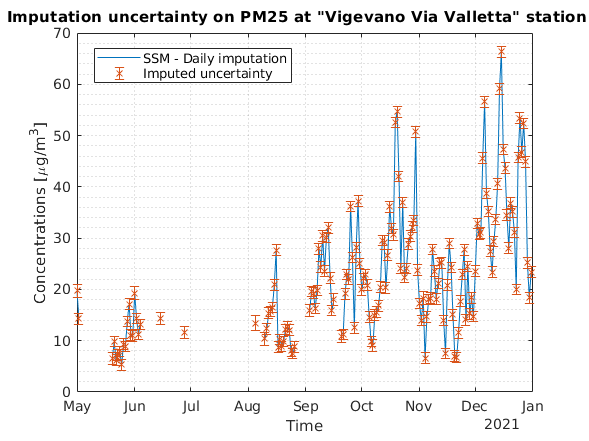}
         \caption{ \centering \label{fig:aq_pm25_unc}}
    \end{subfigure}
\caption{ \textbf{(a)}: Imputed (orange line) and raw (blue dotted line) daily means for the monitoring station named `Vigevano Via Valletta’ (RMSE = $0.3078$). \textbf{(b)}: Daily data with associated imputed uncertainty ($\pm 2\sigma_d$) for the monitoring station named `Vigevano Via Valletta’. \label{fig:aq_pm25_val}}
\end{figure}

Another important experiment concerns the stations that sample pollutants bi-hourly. In this case, if the sampling frequency is regular during the day, we do not expect bias problems in the associated daily time series, although we still have 12 missing values for each day spread every other hour. This situation occurs, for example, in the station called `Vigevano Via Valletta’ In fact, the data on PM$_{2.5}$ concentrations are available with a bi-hourly frequency in the second half of the year 2021. Figure~\ref{fig:aq_pm25} shows the daily mean (blue dotted line), computed without considering missing values, compared to the daily mean computed with our approach (orange line) while Figure~\ref{fig:aq_pm25_unc} shows that the imputed uncertainty, of daily average concentrations computed from imputed hourly time series, is small. 

We examined the dataset for the presence of anomalous values that are clearly outliers after the daily time series construction. Table~\ref{tab:AQ_extreme_val} lists the three instances of anomalous values found. These extreme values have been replaced with the `NaN’ value. It is to be noted that this process is not to be considered a process of searching and removing outliers which is outside the context of this work.

\begin{table}[ht]
\newcolumntype{C}{>{\centering\arraybackslash}X}
\begin{tabularx}{\textwidth}{CCCCCC}
\toprule
\textbf{IDStations}   &  \textbf{Latitude} & \textbf{Longitude} & \textbf{Time} & \textbf{Pollutant} & \textbf{Value} \\
\hline
STA.IT1582A & 45.136 & 8.4452 & 2021-04-27 & AQ\_pm\_10 & 2399 \\
\hline
STA.IT2121A & 45.689 & 8.4584 & 2021-04-04 & AQ\_pm\_25 & 2794 \\
\hline
STA.IT1751A & 44.574 & 8.951 & 2016-10-10 & AQ\_so2 & 152.03 \\
\bottomrule
\end{tabularx}
\caption{Extremely large values for the AQ variables [$\mu g / m^3$] identified in the dataset. The values are replaced with the `NaN’ value. \label{tab:AQ_extreme_val}}
\end{table}

\subsection*{Validation of EM and LI interpolation methods}
\begin{figure}[ht]
\centering
\includegraphics[scale=0.6]{ 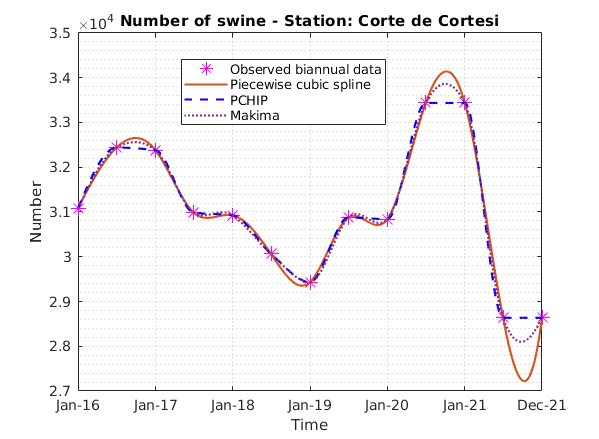}
\caption{Piecewise cubic spline, PCHIP and Makima interpolation methods applied to swine time series for the monitoring station named `Corte de Cortesi’.\label{fig:pchip}}
\end{figure}

The original time resolution for EM and LI variables is lower than daily. In particular, EM variables are provided with a monthly temporal resolution, while LI variables are available every six months. Interpolation techniques are required for daily estimates. To avoid oscillations, overshoots, edge effects and negative values, we use PCHIP~\citep{pchip,PCHIPnumerical} interpolation. This method interpolates the data using a piecewise cubic polynomial while retaining the shape and monotonicity of the original data. Since the points to be interpolated are all positive (e.g. the number of bovines), negative values that could occur with classic splines are avoided. For example, Figure~\ref{fig:pchip} shows the classic piecewise cubic spline, PCHIP and modified Akima piecewise cubic Hermite interpolation (Makima)~\citep{akima}, fitted on data from the `Corte de Cortesi’ station. 

\subsection*{Validation of LA variables}
\begin{figure}[ht]
\centering
\includegraphics[scale=0.5]{ 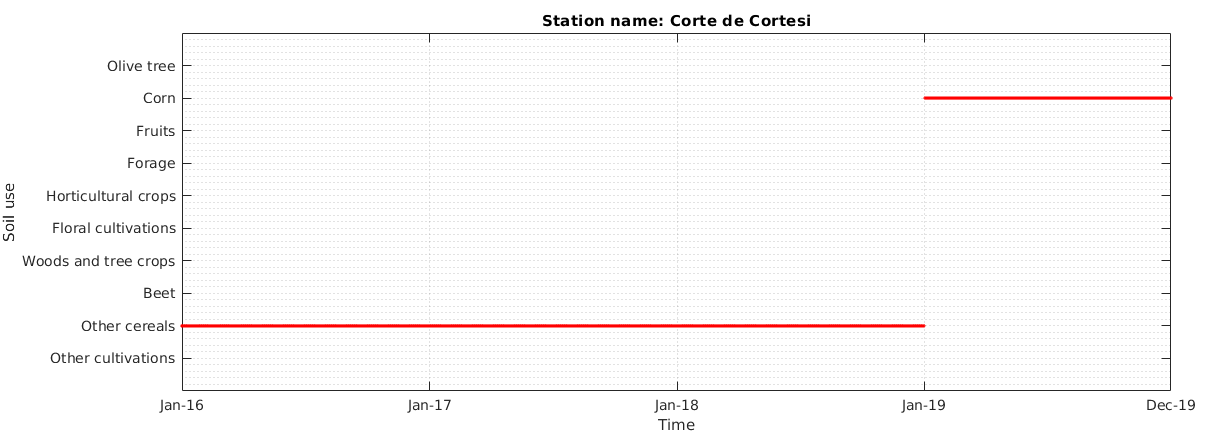}
\caption{Piecewise constant function for soil use provided by SIARL dataset for the station named `Corte de Cortesi’. Note that the SIARL dataset covers data up to 2019 only. The class labels for the SIARL dataset are available in the file named `Metadata\_LA\_SIARL\_labels.csv’ available with the Agrimonia dataset~\citep{Agrimonia_dataset}. \label{fig:la_siarl}}
\end{figure}

Data related to land cover, land and soil use are taken from ERA5-Land, CLC and SIARL datasets, respectively. The main considerations concern land use and soil use. Land use classifies the territory from the urbanisation and/or nature point of view (urban, industrial, road, agricultural, forest, marine and other). Soil use classifies the land from the agricultural production point of view (type of cultivation). Since CLC data are available only for 2018 in the study period (2016-2021), in this work we assume the land use to be constant. This seems appropriate given that the overall area used by various sectors (such as agriculture and infrastructure) changes slowly over time. Also, this assumption is consistent with the fact that the AQ network stations have constant station types. Instead, soil use changes faster because the type of cultivation on the ground can be rotated for greater yield. As an example, Figure~\ref{fig:la_siarl} shows the soil use provided by the SIARL dataset for the station named `Corte dei Cortesi’. The figure shows that the type of cultivation near the station has changed over time.

\section*{Usage notes}

This paper presents an open access spatiotemporal dataset, named Agrimonia dataset \citep{Agrimonia_dataset}, which provides the user with a dataset about ammonia (NH$_{3}$) emissions, agricultural information, air pollution and meteorology in a common spatiotemporal resolution. The dataset is ready to be used `as is’ by those using air quality data for research. The Agrimonia dataset and metadata can be accessed through Zenodo with DOI: \url{https://doi.org/10.5281/zenodo.6620530}. In the same repository, metadata and supplementary materials are provided to better understand the dataset. As previously mentioned, this dataset was initially compiled for the \emph{AgrImOnIA project} and will be updated when new variables will be available.

\section*{Code availability}

To replicate our work, downloadable helper functions can be used to filter and merge data from different sources. The code used to extract and process all the datasets are developed using Matlab and R. The codes are available at \url{https://github.com/Agrimonia-project/Agrimonia_Data.git}, with the user instructions included in the respective `README.md’ files.

\section*{Acknowledgements}
This research was funded by Fondazione Cariplo under the grant 2020-4066 ``AgrImOnIA: the impact of agriculture on air quality and the COVID-19 pandemic'' from the ``Data Science for science and society'' program. The authors would like to thank ARPA Lombardia for making the AQ data available and for their fundamental support.

\section*{Author contributions statement}
A.F. is the PI of the AgrImOnIA project and supervised all the phases of dataset construction and paper writing. J.R. wrote the initial draft of the manuscript and was in charge of the figures. J.R., P.M., A.F.M. and Q.S. designed the dataset and implemented the data extraction, cleaning, harmonisation and imputation code. In particular, J.R. and P.M. led the AQ and LI data handling, A.F.M. lead the WE and EM data handling, and Q.S. lead the LA data handling. A.F., P.M., M.C., F.F., N.G., R.I. and P.O. participated and provided input during all stages of data harmonisation and paper authoring. They also performed the data quality check.

\section*{Competing interests} 
The authors declare no competing interests.

\end{document}